\documentclass[conference]{IEEEtran}
\IEEEoverridecommandlockouts
\usepackage{cite}
\usepackage{amsmath,amssymb,amsfonts}
\usepackage{algorithmic}
\usepackage{graphicx}
\usepackage{textcomp}
\usepackage{xcolor}
\def\BibTeX{{\rm B\kern-.05em{\sc i\kern-.025em b}\kern-.08em
    T\kern-.1667em\lower.7ex\hbox{E}\kern-.125emX}}
    
\renewcommand{\vec}[1]{\mathbf{#1}}
\newcommand{\mtx}[1]{\mathbf{#1}}
    
\begin{document}

\title{Generalized Noncoherent Space-Time Block Codes from Quantum Error Correction\\
\thanks{This  work  was  supported  by  the  U.S.  Office  of  Naval Research  under  Grant  No.  N00014-17-1-2107.}
}

\author{\IEEEauthorblockN{S. Andrew Lanham\IEEEauthorrefmark{2}, Eli Bradley\IEEEauthorrefmark{1}, Brian R. La Cour \IEEEauthorrefmark{2}}
\IEEEauthorblockA{\IEEEauthorrefmark{1}Department of Computer Science} \IEEEauthorblockA{\IEEEauthorrefmark{2}Applied Research Laboratories} \IEEEauthorblockA{The University of Texas at Austin, Austin, TX 78758 USA\\
Emails: sa\_lanham@utexas.edu, elibradley@utexas.edu, blacour@arlut.utexas.edu}
}

\maketitle

\begin{abstract}
This paper generalizes results in noncoherent space-time block code (STBC) design  based on quantum error correction (QEC) to new antenna configurations. Previous work proposed QEC-inspired STBCs for antenna geometries where the number of transmit and receive antennas were equal and a power of two. In this work we extend these results by providing QEC-inspired STBCs applicable to all square antenna geometries and some rectangular geometries where the number of receive antennas is greater than the number of transmit antennas. We derive the maximum-likelihood decoding rule for this family of codes for the special case of Rayleigh fading with additive white Gaussian noise. We present Monte Carlo simulations of the performance of the codes in this environment for a three-antenna square geometry and a three-by-six rectangular geometry. We demonstrate competitive performance for these codes with respect to a popular noncoherent differential code.
\end{abstract}


\section{Introduction}
Noncoherent communication refers to a particular method of wireless communication wherein neither the transmitter nor the receiver structures form an instantaneous estimate of the communication channel. While noncoherent communication often incurs a performance penalty against methods employing channel estimation, it remains useful in certain situations, such as when there is insufficient time to form accurate estimates or in settings where the communication equipment is tightly resource constrained \cite{diggavi2004great}. Noncoherent coding techniques have been proposed to mitigate the constraints imposed by channel estimation, such as through differential signaling \cite{hughes2000differential,tarokh1999differential} or more general space-time block codes \cite{jing2003unitary,lanham2019noncoherent}. 

Notably, noncoherent designs have a long history of being integrated into multiantenna topologies \cite{hochwald2000unitary,zheng2002communication}. Multiple antennas allow code designs to continue to exploit spatial diversity, often through space-time coding. Spatial diversity, along with noncoherent design, is a crucial resource to consider in highly resource-constrained settings. In particular, blocklength-constrained or low-latency environments paired with slowly varying channels can render a system susceptible to channel outages. In these situations spatial diversity creates opportunities to leverage additional independent fading paths to reduce the overall bit error probability of the system \cite{zheng2003diversity}. Consequently space-time codes and other methods employing spatial diversity have continuing relevance for ultra-reliable and resource-constrained settings \cite{durisi2015short}. 

With this motivation, a need arises to support noncoherent communication over a variety of antenna geometries and with high diversity. Flexible noncoherent space-time code designs allow for integration with a wide variety of transceiver characteristics. Existing designs, however, are significantly topologically constrained. For example, differential designs are based on specific matrix groups whose generalizations are not always clear. Similarly, the recent quantum-inspired designs on which this work is based apply only for square power-of-two antenna geometries \cite{cuvelier2021quantum}. Furthermore, with diversity-maximizing designs it is crucial to support as many small-antenna configurations as possible, as decoding overhead in diversity-maximizing approaches grows prohibitively large with more antennas \cite{diggavi2004great}. Generalizations to new antenna geometries are significant in this light.  

To this end, this work significantly generalizes a recently proposed family of codes based on quantum error correction \cite{lanham2019noncoherent,cuvelier2021quantum}. This work adopts the framework presented in \cite{cuvelier2021quantum} for building quantum-inspired noncoherent space-time block codes and extends it by proposing a novel stabilizer group supporting noncoherent communication for all square antenna geometries and select geometries for which the number of receive antennas is greater than the number of transmit antennas. The proposed codes are shown to have maximal diversity and similar properties to their predecessors for square, power-of-two geometries. We explore the performance of the codes in a canonical Rayleigh-fading setting with additive white Gaussian noise. We compare their performance to a widely used noncoherent code, the differential STBC \cite{hughes2000differential}. The results demonstrate competitive performance and point to the utility of these codes in constrained settings. 

The possibilities for antenna configurations and code lengths are best understood by delineating the compatible configurations for which a code may be defined. For comprehension, we propose the tuple of parameters $(M, N, T, d)$ as a means of describing the family of codes. Define a \textit{configuration} as a tuple of positive integer parameters $(M, N, T, d)$. The quantum-inspired STBCs of this work allow for recovery of a vector in $\mathbb{C}^d$ with full diversity, meaning, informally, that the encoded vector passes through every available independent fading path. Thus, a code with parameters $(M, N, T, d)$ is designed for $M$ transmit antennas, $N$ receive antennas, a coherence time of $T$, during which the fading channel is assumed to be static. Finally the parameter $d$ refers to the supported dimension of the encoded vector. We continue the analysis with a review of background concepts in Section \ref{sec:bckgrnd} before continuing to the system model in Section \ref{sec:sys_mod}. In Section \ref{sec:square} we describe the first family of codes, which pertain to square antenna geometries 
%
%
while in Subsection \ref{subsec:rect} we cover the rectangular cases
%
%
before concluding remarks in Section \ref{sec:conc}. 

\subsection{Notation}
We use bold lowercase letters such as $\vec{a}$ to denote column vectors, and bold uppercase letters such as $\mtx{A}$ to denote matrices. We use non-bold italics letters such as $a$ to denote scalars. We denote the element in the $i^{\rm th}$ row and $k^{\rm th}$ column of a matrix $\mtx{A}$ by $[\mtx{A}]_{i,k}$. In general we denote the $k\times k$ identity matrix by $\mtx{I}_k$. For a square matrix $\mtx{A}$ we use $\mathrm{Tr}(\mtx{A})$ to denote its trace, $\mathrm{det}(\mtx{A})$ its determinant, $\mtx{A}^\mathsf{T}$ its transpose, and $\mtx{A}^*$ its conjugate transpose. We use $\mathbb{E}[\cdot]$ to denote expectation and $\Pr[\cdot]$ to denote probability. We use
$\otimes$ to denote the tensor product when acting on vector spaces (i.e. $\mathbb{C}^2\otimes\mathbb{C}^2$) and to denote the Kronecker product when acting on vectors or matrices. We use $\mathcal{CN}(\boldsymbol{\mu},\boldsymbol{\Sigma})$ to denote a complex circularly symmetric normal distribution with mean $\vec{\mu}$ and covariance $\mtx{\Sigma}$. If $a = n b$ for $a,b \in \mathbb{N}$ and some $n \in \mathbb{N}$, we write $b | a$. We abbreviate the set of integers $\{0, \ldots, d-1\}$ as $[d]$. Finally, when the dimension $d$ is apparent, we use $\vec{e}_i$ as the $i^{\textrm{th}}$ standard basis vector of dimension $d$. 

\section{Mathematical Background} 
\label{sec:bckgrnd}

Stabilizer codes are constructions from quantum error correction that are used to protect quantum information against the deleterious effects of the environment. Their relevance to noncoherent space-time block coding was previously established in Refs.\ \cite{lanham2019noncoherent} and \cite{cuvelier2021quantum}, where the protective effects of the code can also be used to preserve information against fading without channel estimates. Early quantum stabilizer codes were based on the $2\times 2$ Pauli matrix group and thus protect quantum systems comprised of two-dimensional quantum systems, or qubits \cite{gottesman1997stabilizer}. Because the construction uses qubits as the basic unit of quantum information, the dimension of the encoded vector is limited to a power of two, and this is a limitation appearing in prior work \cite{lanham2019noncoherent,cuvelier2021quantum}. In this section we review crucial mathematical details that allow for a generalization of the stabilizer code construction to new antenna configurations. These constructions are loosely based on a generalization of quantum stabilizer codes for systems of arbitrary, but finite dimension \cite{gottesman2001encoding}. 

To generalize the stabilizer code construction to arbitrary dimensions we seek a framework for encoding a $d$-dimensional subsystem, with $d \ge 2$, into a $K$-dimensional encoded system, where $d | K$. The desired stabilizer code preserves the configuration of a $d$-dimensional vector against the effects of a set of linear operators, referred to henceforth as \textit{error operators}. We express error operators in terms of an \textit{error basis}, a matrix basis spanning the space of linear operators acting on a $d$-dimensional system. The generalized error basis $\mathcal{P}_d$ is a set of $d^2$ matrices defined as 
\begin{equation}
    \mathcal{P}_d = \{ \mtx{X}_d^a \mtx{Z}_d^b : a, b \in [d] \}
\end{equation}
where, for $i \in [d]$ and $\omega = e^{j2\pi/d}$,
\begin{equation}
    \label{eq:shift_op}
    \mtx{X}_d \vec{e}_i = \vec{e}_{i+1}
\end{equation}
and
\begin{equation}
    \label{eq:clock_op} 
    \mtx{Z}_d \vec{e}_i = \omega^{i} \vec{e}_i
\end{equation}
for an arbitrary standard basis vector $\vec{e}_i$. The matrices $\mtx{X}_d$ and $\mtx{Z}_d$ are often called \textit{shift} and \textit{clock} operators, respectively, and satisfy the relation 
\begin{equation}
    \mtx{Z}_d \mtx{X}_d = \omega \mtx{X}_d \mtx{Z}_d \, ;
\end{equation}
that is, they commute up to a phase factor. The set $\mathcal{P}_d$ generalizes the elements of the Pauli group in the following ways. It forms a basis of unitary matrices for the set of linear operators on $\mathbb{C}^d$, and it is also a group under standard matrix multiplication \cite{gottesman2001encoding}. For $K = d^n$, the set of errors in $\mathbb{C}^K = \mathbb{C}^d \otimes \mathbb{C}^d \otimes \ldots \otimes \mathbb{C}^d $ is spanned by the set of all possible Kronecker products of length $n$ between elements of $\mathcal{P}_{d}$.

The theory of stabilizer codes and the ensuing quantum-inspired generalized space-time block code require a suitably defined \textit{stabilizer group} with respect to the error basis $\mathcal{P}_d$ \cite{lanham2019noncoherent,Mike&Ike}. The stabilizer group $S$ is a commutative subgroup of $d\times d$ matrices under multiplication that does not contain the phase-shifted identity $\omega \mtx{I}_d$. We say that a subspace of a vector space is \textit{stabilized} by $S$ if for any vector $\vec{v}$ in the subspace and $\mtx{F} \in S$, we have $\mtx{F} \vec{v} = \vec{v}$. In the code construction, the stabilizer group generates a \textit{codespace}, defined as a subspace stabilized by $S$. The requirement that no phase-shifted identity be in the stabilizer group arises because such elements only stabilize the trivial subspace. In the wireless setting, the codespace generated by the stabilizer group preserves information against fading if each error in the basis has a unique commutation relationship with respect to the stabilizer group generators \cite{cuvelier2021quantum}. The stabilizer group generators and their commutation properties, therefore, play a role similar to syndromes in classical linear error correcting codes. 

For the remainder of this manuscript, we propose and analyze a set of stabilizer group generators based on the Pauli clock and shift matrices for arbitrary finite dimension. We will verify the properties of the stabilizer generators with respect to the error basis and show that the resulting codespace can be used to encode information for noncoherent communication across a wireless link. In particular, we prove that the generators commute, which is both necessary and sufficient to verify that the stabilizer group is commutative, and we show that each error generates a unique set of commutation relationships with the stabilizer generators, henceforth referred to as \textit{syndromes} for that error.

\section{System Model}
\label{sec:sys_mod}
The system model we adopt for the wireless communication environment is similar to previous work on this subject \cite{cuvelier2021quantum,lanham2019noncoherent}. Following the canonical models for wireless noncoherent communication \cite{hochwald2000unitary,zheng2002communication}, we assume that there are $M$ antennas employed at the transmitter and $N$ antennas employed at the receiver. We model the channel as a single-tap coefficient matrix $\mtx{H} \in \mathbb{C}^{N \times M}$ capturing a narrowband environment and assume there is additive noise $\mtx{N}$ at each time instant and receive antenna. Finally, we assume that the fading matrix $\mtx{H} $ is static for multiple discrete coherence intervals of length $T$ before changing to a new, independent configuration. The parameter $T$ is the \textit{coherence time} of the channel. In Section \ref{sec:decoder}, we make more specific assumptions about the fading and additive noise models in order to derive a maximum likelihood decoder for a common fading environment, but those additional assumptions are not integral to the code design. In particular, distributional information about the fading matrix and noise is not required for successful code design. The code is designed to transmit information without error in the infinite SNR environment, regardless of fading, as long as $\Pr[\mtx{H} = \mtx{0}] =0$. The model for the received signal over the coherence interval is thus given as 
\begin{equation}
    \mtx{Y} = \mtx{HX} + \mtx{N} \, , 
\end{equation}
with $\mtx{X}\in \mathbb{C}^{M \times T }$ the set of transmitted symbols, and $\mtx{Y} \in \mathbb{C}^{N \times T}$ the received symbols. 

Following previous work, we apply a standard vectorization identity to the channel model before introducing the stabilizer formalism for code design \cite{lanham2019noncoherent}. The vectorized channel is given as 
\begin{equation}
    \label{eq:system_model_vectorized}
    \vec{y} = \overline{\mtx{H}} \vec{x} + \vec{n} 
\end{equation}
where $\vec{y} = \textrm{vec}(\mtx{Y})$, $\vec{x} = \textrm{vec}(\mtx{X})$ and $\vec{n} = \textrm{vec}(\mtx{N})$ are column vectors formed by stacking the columns of the corresponding matrices and $\overline{\mtx{H}} = \mtx{I}_T \otimes \mtx{H}$ is block diagonal. We also assume a unity average power in the transmission. This is enforced by ensuring that each codeword have equal average power, meaning that for $\vec{x} \in \mathcal{C} $ our codebook design will have 
\begin{equation}
    \label{eq:avg_power}
    \vec{x}^* \vec{x} = T \, .
\end{equation} 
This also ensures that the time-averaged power is equal to $1$ i.e. for a codebook $\mathcal{C}$, we have $\mathbb{E}_{\mathcal{C}}[\mathbf{x}^* \mathbf{x}]/T = 1$

The system model presented introduces the opportunity to define codes for a variety of antenna configurations, coherence times, and dimension $d$. A dimension of $d$ implies that $d$ symbols, or a vector in $\mathbb{C}^d$, is encoded and protected against the effects of the channel in a noncoherent setting. There are $MN$ fading paths, so transmission of a $d$-dimensional vector requires at least $MNd$ input degrees of freedom. The available input degrees of freedom are $MT$, introducing the requirement that $MT \geq MNd$; i.e., 
\begin{equation}
    T \geq Nd  \, . 
\end{equation}
Another requirement is that $d \geq 2$, as the noncoherent coding presented requires phase insensitivity, which is not achievable with unit-power scalar-valued quantities. A final requirement is that $N \geq M$; codes may not be constructed for the case $N  < M$, as is shown in Section \ref{subsec:rect}. 

We divide our presentation into two naturally different antenna configurations, in particular, the $M = N$ case and $M < N$ case. We also restrict our focus to instances for which $N | T$; i.e. the coherence interval is a multiple of the number of receive antennas. This ensures that the dimension, $d$, is an integer. In Section \ref{sec:square} we consider square configurations
\begin{equation}
\mathcal{F}_1 = \{ (M, N, T, d) : M = N, T \geq Nd, N | T, d \geq 2  \} ,
\end{equation}
while in Section \ref{subsec:rect} we consider rectangular configurations of the form
\begin{equation}
\mathcal{F}_2 = \{ (M, N, T, d) : M | N, T \geq Nd, N | T, d \geq 2  \} .
\end{equation}

\section{Square Configurations} 
\label{sec:square} 

For the square configuration, we assume $N$ transmit and receive antennas. The motivation for the use of quantum codes follows from a particular decomposition of the channel, presented first in \cite{lanham2019noncoherent} and extended here with the generalized Pauli matrices. In particular, the coherence time of $T$ allows for the decomposition
\begin{equation}
    \label{eq:matrix_expansion}
    \overline{\mtx{H}} = \sum_{i=0}^{N^2-1} c_i \bigl( \mtx{I}_T \otimes \mtx{B}_i \bigr) 
\end{equation}
where $\mtx{B}_i$ are the elements of the $N$-dimensional generalized Pauli basis $\mathcal{P}_N$. The basis expansion allows for a characterization of the coefficients as $c_i = \mathrm{Tr}(\mtx{B}_i^* \mtx{H})/N$, where the scalar normalization factor emerges from the fact that the basis matrices have $\mathrm{Tr}[(\mtx{I}_T \otimes \mtx{B}_i)^* (\mtx{I}_T \otimes \mtx{B}_i)] = NT$ \cite{cuvelier2021quantum}. The square antenna configurations thus introduce a basis of $N^2$ error operators, corresponding also to the number of diversity branches in the system. The stabilizer code construction must then follow from identifying stabilizer generators possessing unique commutation relationships with the error basis $\mathcal{E}_{N,T}$ which we define as the set of $N^2$ operators of the form $\mtx{E}_i = \mtx{I}_T \otimes \mtx{B}_i$. 


We now delineate the construction of the corresponding stabilizer group, $\mathcal{S}_{N,T}$. As with any stabilizer construction, in order to generate a nontrivial codespace the stabilizer group must not contain a phase shifted identity. A set of stabilizer generators possessing unique commutation properties with error operators $\mathcal{E}_{N,T}$ is 
\begin{equation}
    \begin{split}
    \label{eq:stab_group}
    \mathcal{S}_{N,T} = \{ &\mtx{I}_{T/N} \otimes \mtx{X}_N \otimes \mtx{X}_N , \;\; \mtx{I}_{T/N} \otimes \mtx{Z}_N \otimes \mtx{Z}_N^{-1}  \} \, .
    \end{split}
\end{equation}
We first prove that the generators commute, which is sufficient to show that the group commutes. It is easy to see from the distributive property of the Kronecker product and the commutation relationships of the generalized Pauli basis elements that the two generators commute. In particular $\mtx{X}_N \mtx{Z}_N = \omega \mtx{Z}_N \mtx{X}_N $ and $\mtx{X}_N \mtx{Z}^{-1}_N = \omega^{-1} \mtx{Z}_N^{-1} \mtx{X}_N$, while identity elements trivially commute. This leaves a total commutation factor of $\omega \omega^{-1} = 1$ for the two generators, and thus they commute. 

To show that the stabilizer generators generate distinct error syndromes with respect to the $N^2$ error operators, we restrict the focus to the rightmost $N$-dimensional operator of the stabilizer generators. The commutation relationship between members of $\mathcal{E}_{N,T}$ and the two generators in $\mathcal{S}_{N,T}$ are determined by these rightmost operators. For error operators, we use a shorthand representation where the error set is described as $\mathcal{E} = \{ \mtx{I}_T \otimes \mtx{E}_{(a,b)} : \mtx{E}_{(a,b)} = \mtx{X}_N^a \mtx{Z}_N^b, \; a,b \in [N] \}$. Focusing on the relations between rightmost operators, we can see firstly that 
\begin{equation}
    \begin{split}
    \mtx{E}_{(a,b)} \mtx{X}_N &= \mtx{X}_N^a \mtx{Z}_N^b  \mtx{X}_N \\
    &=  \mtx{X}_N^a \omega^b \mtx{X}_N \mtx{Z}_N^b \\ 
    &=  \omega^{b} \mtx{X}_N \mtx{E}_{(a,b)} \, .
    \end{split}
\end{equation}
Through this commutation relationship, the first stabilizer generator produces a unique signature of the shift operator degree parameter $b$ and provides distinguishability between shift parameters. In a similar manner we have for the second stabilizer generator that 
\begin{equation}
\begin{split}
    \mtx{E}_{(a,b)} \mtx{Z}^{-1}_N &= \mtx{X}_N^a \mtx{Z}_N^{-1} \mtx{Z}_N^b \\
    &=  \omega^{-a} \mtx{Z}_N^{-1} \mtx{X}_N^a \mtx{Z}_N^b \\ 
    &=  \omega^{-a} \mtx{Z}^{-1} \mtx{E}_{(a,b)} \, .
\end{split}
\end{equation}
We can consequently deduce the degree of the shift and clock operators arising in each of the errors. Each particular pair of degrees is unique to one error. 

The unique coefficients arising in the respective commutation relationships imply uniqueness of commutation relationships. Via results in stabilizer coding \cite[Section II-D]{cuvelier2021quantum} this implies that a $d$-dimensional codespace can be identified such that each error operator in the error basis expansion maps the codespace to a separate, orthogonal $d$-dimensional subspace of the system space. In particular, the generators, because of their unique commutation properties, identify the codespace for the space-time code as the mutual $+1$ eigenspace of the stabilizer generators. This is identified by its projector 
\begin{equation}
    \label{eq:codespace_nonsq}
    \mtx{P}_0 = \frac{(\mtx{I}_{NT} + \mtx{S}_1 )}{2} \frac{(\mtx{I}_{NT} + \mtx{S}_2 )}{2} \, . 
\end{equation}
By applying the appropriate commutation coefficients $z=(z_1,z_2)$, we can identify the subspaces associated with the syndromes. We have
\begin{equation}
    \label{eq:subspace_proj}
    \mtx{P}_z = \frac{(\mtx{I}_{NT} + z_1\mtx{S}_1 )}{2} \frac{(\mtx{I}_{NT} + z_2 \mtx{S}_2 )}{2} ,  
\end{equation}
where $z_1$ and $z_2$ represent the desired commutation coefficient, i.e. $\omega^{-a}$ or $\omega^{b}$ depending on the desired error subspace. In the wireless communication setting, each of these subspaces encode a realization of the encoded vector transmitted along a unique diversity path. The subspaces, as identified by the projector, have a bijective relation to the errors $\mtx{E}_{(a,b)}$. The codespace thus encodes a vector $\vec{s} \in \mathbb{C}^d$ and introduces maximum diversity $N^2$ by allowing recovery of the encoded vector (up to a global phase), after independent fading, in subspaces identified by the error operators. 

Encoding is achieved with code matrix $\mtx{C}$, which has as its columns the unit vectors spanning the mutual $+1$ eigenspace of the stabilizer generators. The eigenspace has dimension $d$ with proper construction \cite{cuvelier2021quantum}. The encoded symbol $\vec{s}$ can be any unit vector in $\mathbb{C}^d$, and will be chosen from a codebook $\mathcal{C}$. The encoded transmission is thus $\vec{x} = \sqrt{T} \mtx{C} \vec{s}$, and gives rise to an updated model from Equation (\ref{eq:system_model_vectorized}), where now 
\begin{equation}
    \label{eq:received_encoded}
    \vec{y} = \sqrt{T} \overline{\mtx{H}} \mtx{C} \vec{s} + \vec{n} \, . 
\end{equation}
Note that with $\mtx{C}$ and $\vec{s}$ selected to be of unit energy, i.e. $\mtx{C}^* \mtx{C} = \mtx{I}_d$ and $\vec{s}^*\vec{s} = 1$, the time-averaged radiated power is unity. Details of codebook design are considered in Section \ref{sec:decoder} in light of the derived maximum-likelihood detection strategy. 

We now prove that there exist recovery operators based on the projectors $\mtx{P}_z$ and error operators $\mtx{B} \in \mathcal{P}_N$. Because the stabilizer group proposed in (\ref{eq:stab_group}) produces unique syndromes, it follows \cite{cuvelier2021quantum} that we can produce a recovery operator $\mtx{R}_z = \mtx{E}_z^* \mtx{P}_z$ for each syndrome $z$, where $z = (a,b)$. The projectors $\mtx{P}_z$ are mutually orthogonal, and so each recovery operator may be applied separately to obtain a set of recovered signals. Application of each projector $\mtx{P}_z$ recovers an independently faded encoded vector associated with each error. That is,  
\begin{subequations}
    \begin{alignat}{2}
    \label{eq:recovery_op_1}
    \vec{q}_z &=  \mtx{E}_z^* \mtx{P}_z \bigl( \sqrt{T} \overline{\mtx{H}} \mtx{C} \vec{s} + \vec{n}) \\ 
    &= \sqrt{T} \mtx{E}_z^* \mtx{P}_z \sum_j c_j \mtx{E}_j \mtx{C} \vec{s} + \vec{n}_z \\ 
    \label{eq:recovery_op_3}
    &= \sqrt{T} c_z \mtx{C} \vec{s} + \vec{n}_z
    \end{alignat}
\end{subequations}
where recovery is enabled by the fact that $\mtx{P}_z \sum_j c_j \mtx{E}_j = c_z \mtx{E}_z$. The indexing of the error operators $\mtx{E}_z, z = (a,b)$ is made possible using the bijection between the $N^2$ error operators and the $N^2$ orthogonal subspaces created by the stabilizer code construction. The projected noise vectors $\vec{n}_z = \mtx{E}^*_z\mtx{P}_z \vec{n}$ with $\vec{n} \sim \mathcal{CN}(\vec{0}_{NT}, \sigma^2 \mtx{P}_z )$ remain IID and become $\mathcal{CN}(\vec{0}_{NT}, \sigma^2\mtx{P}_0)$ after rotation back to the codespace by $\mtx{E}_z^*$. The coefficients $c_z$ are simply a re-indexing of the coefficients appearing in the basis expansion.


\subsection{Non-Square Configurations}
\label{subsec:rect}
For the purposes of code construction, non-square configurations are limited to instances for which $M < N$, i.e. when there are more receive antennas than transmit. This follows from a simple dimensional analysis and proof by contradiction: Assume that $M > N$. The channel matrix $\mtx{H}\in \mathbb{C}^{N \times M}$ is spanned by a set error operators for that dimension which we abbreviate as $\mathcal{E}_{(N,M)}$. We have $\textrm{rank}(\mtx{E}_i) \leq N$ for $\mathbf{E}_i \in \mathcal{E}_{(N,M)}$, and the rank of the full error operators $\mtx{I}_T \otimes \mtx{E}_i \leq NT$. Define the recovery operator $\mtx{F}_i$ to be the inverse of $\mtx{E}_i$. Then we require $\mtx{F}_i \mtx{E}_i = \mtx{I}_{MT}$ and necessarily $\textrm{rank}(\mtx{F}_i \mtx{E}_i) = MT$. This gives $\textrm{rank}(\mtx{F}_i \mtx{E}_i) \leq \textrm{rank}(\mtx{I}_T \otimes \mtx{E}_i)$. Thus it follows that $MT \leq NT$ and thus $M\leq N$. 
 
Processing for non-square configurations proceeds from imposing that $M | N$, i.e. the number of receive antennas is a multiple of the number of the transmit antennas. This effectively allows each receive antenna block of size $M$ to be treated as an independent block, and each such block can be processed separately with a stabilizer code defined for $M\times M$ structure. We spend the remainder of the section detailing this protocol. First, we define the non-square generalized Pauli group for the case $M | N$. The $MN$ basis matrices are defined as block matrices composed of the structures outlined in Equations (\ref{eq:shift_op}) and (\ref{eq:clock_op}). We abbreviate the generalized Pauli matrices $ \mtx{R}(a,b) = \mtx{X}^a\mtx{Z}^b$ where each $\mtx{R}(a,b) \in \mathcal{P}_M$. The relevant block structures for the $M \times N$ antenna configuration are 
\begin{equation}
    \label{eq:basis_nonsq}
    \begin{split}
    &\mathcal{P}_{M,N} = \\
    &\left\{ 
    \begin{bmatrix}
    \mtx{R}(a,b) \\
    \vec{0}_{M\times M} \\ 
    \vdots  \\ 
    \vec{0}_{M\times M}
    \end{bmatrix} 
    : a,b \in [d] 
    \right\} \cup ... \cup 
    \left\{ 
    \begin{bmatrix}
    \vec{0}_{M\times M} \\
    \vec{0}_{M\times M} \\ 
    \vdots  \\ 
    \mtx{R}(a,b)
    \end{bmatrix} 
    : a,b \in [d] 
    \right\}
    \end{split} 
\end{equation}
The union of these sets enumerates a family of $\ell M^2 = MN$ matrices, where $\ell M = N$. Define $\mtx{B}(a,b,l)$ to be the basis element $\mtx{R}(a,b)$ corresponding to block $l \in [\ell]$. Note that each $\mtx{B}(a,b,l)$ is semi-unitary and that the set of operators in Equation (\ref{eq:basis_nonsq}) forms a basis for the operator space $\mathbb{C}^{N \times M}$. 

The vectorized signal model for non-square configurations is similar to Equation (\ref{eq:system_model_vectorized}) but now with $\mtx{X} \in \mathbb{C}^{MT \times 1}$, $\vec{y} \in \mathbb{C}^{NT \times 1}$, $\mtx{H} \in \mathbb{C}^{N \times M}$, and $\overline{\mtx{H}} = \mtx{I}_T \otimes \mtx{H}$. The set of matrices $\mathcal{P}_{M,N} = \{ \mtx{B}(a,b,l) : a,b \in [M], l \in [\ell]\}$ form the basis for channel matrix $\mtx{H}$. The error operators for the vectorized channel are simply $\mathcal{E}_{(M,N,T)} = \{  \mtx{I}_T \otimes \mtx{B}(a, b,l) : \mtx{B}(a,b,l) \in \mathcal{P}_{M,N} \} $. Expanding the vectorized channel operator using the basis $\mtx{B}(a, b, l)$ gives 
\begin{equation}
    \overline{\mtx{H}} = \mtx{I}_T \otimes \sum_{l = 0}^{\ell-1} \sum_{a = 0}^{M-1} \sum_{b = 0}^{M-1} c(a,b,l)\mtx{B}(a, b,l)  
\end{equation}
where $c(a,b,l) = \textrm{Tr}(\mtx{B}(a, b,l)^* \mtx{H})/N$. The received signal model can be written in a block structure to demonstrate the possibility for independent processing. That is,  
\begin{equation}
    \vec{y} = 
    \textrm{vec}\bigl( 
    \begin{bmatrix}
    \vec{y}_{0,1} \ldots \vec{y}_{\ell-1,1}, \ldots ,\vec{y}_{0,T} \ldots , \vec{y}_{\ell-1, T}  
    \end{bmatrix}\bigr) + \vec{n} 
\end{equation}
where we define $\vec{y}_{i,t}$ to be the length-$M$ received vector for the $l^{\textrm{th}}$ block of receive antennas and the $t^{\textrm{th}}$ timeslot.

This framework can be integrated with the stabilizer construction, which ordinarily only applies to square configurations, by treating decoding as a block operation on $\ell$ separate sets of $M \times M$ structures. Besides this distinction in decoding, the stabilizer encoding process is identical. In particular, since each $M\times M$ sub-block of the error matrices $\mathcal{P}_{M,N}$ is a square matrix, and because each $M\times M$ block can be processed separately, it is sufficient to confirm relationships between the $M \times M$ error operators $\mtx{R}(a,b)$ and a stabilizer construction of choice. In effect, a satisfying stabilizer construction turns out to be identical to Equation (\ref{eq:stab_group}). The projector into the code space is thus derived identically as Equation (\ref{eq:codespace_nonsq}). 

\section{Maximum Likelihood Decoding (ML) and Performance}
\label{sec:decoder}
The maximum likelihood detection strategy for square antenna configurations begins with the received signal model of Equation (\ref{eq:received_encoded}) and broadly follows the developments of \cite{cuvelier2021quantum}, although the processing described in this manuscript pertains to a broader class of antenna configurations. We review and detail crucial distinctions in the decoder design in this section. Firstly, for the decoder design, we adopt a standard assumption of Rayleigh fading in white Gaussian noise. This environment is modeled by assuming entries $[\overline{\mtx{H}}]_{i,j} \sim \mathcal{CN}(0,1)$ i.i.d. and $\vec{n} \sim \mathcal{CN}(\vec{0}_{NT},\sigma^2 \mtx{I}_{NT})$. 

The stabilizer code allows for recovery of $N^2$ independently faded encoded vectors as demonstrated in Equations (\ref{eq:recovery_op_1})-(\ref{eq:recovery_op_3}). We demonstrate the compatibility of this approach with the maximum-likelihood detection rule for the signal model of Equation (\ref{eq:received_encoded}). The basis decomposition of $\overline{\mtx{H}}$ presented in Equation (\ref{eq:matrix_expansion}) expresses the channel matrix as a linear combination of orthogonal basis matrices. The associated coefficients $c_i$ are thus projections of the jointly-Gaussian channel matrix $\mtx{H}$ onto the orthogonal basis. The $c_i$ are also jointly Gaussian, therefore. The received signal model can thus be expressed using this decomposition to yield
\begin{equation}
    \vec{y} = \sqrt{T} \sum_{j = 0}^{|\mathcal{P}_N| - 1} c_j \mtx{E}_j \mtx{C} \vec{s} + \vec{n}
\end{equation}
with the associated maximum-likelihood decoding problem
\begin{equation}
    \hat{\vec{s}} = \arg\max_{\vec{s} \in \mathcal{C}} f_{\vec{y}| \vec{s}} (\vec{y}| \vec{s}) \, . 
\end{equation}
By assumption we have independence between the channel and the additive noise. We may use completeness and orthogonality of the set of projectors of Equation (\ref{eq:subspace_proj}) to conclude that the set of signals $\{ \mtx{P}_z \vec{y}, \; z \in [N^2] \} $ are a sufficient statistic for $\vec{y}$. Furthermore, because each $\mtx{E}_z^*$ effects a unitary transformation on these signals, we have sufficiency of the signals $\vec{q}_z$ in Equation (\ref{eq:recovery_op_3}). 

The vectors $\vec{q}_z$ may be further decoded by applying the inverse of the encoding operator $\mtx{C}$, $\mtx{C}^*$. We may thus define 
\begin{subequations}
    \begin{alignat}{2}
    \label{eq:decoded_signal}
    \vec{p}_z &= \mtx{C}^* \vec{q}_z \\ 
    &= \mtx{C}^* (c_z \sqrt{T} \mtx{C} \vec{s} + \vec{n}_z )
    \end{alignat} 
\end{subequations}
Let $\tilde{\vec{n}}_z = \frac{1}{\sqrt{T}} \vec{n}_z$ and $\tilde{c}_z = \frac{1}{\sqrt{T}} c_z$. This yields the simplified received model
\begin{equation}
\vec{p}_z = \tilde{c}_z \vec{s} + \tilde{\vec{n}}_z
\end{equation}
where, using $T = dN $ we have $\tilde{\vec{n}}_z \sim \mathcal{CN}(0, \frac{\sigma^2}{d} \mtx{I}_d )$ and $\tilde{c}_z \sim \mathcal{CN}(0, 1)$. Following the derivation in \cite[Sec. V]{cuvelier2021quantum} with the slightly different normalization factors, the maximum-likelihood detection strategy is identically 
\begin{equation}
    \hat{\vec{s}}_{\textrm{ML, sq}} = \textrm{arg}\max_{\vec{s} \in \mathcal{C}} \vec{s}^* \sum_{n = 0}^{N^2 - 1} \vec{p}_n \vec{p}_n^* \vec{s} \, . 
\end{equation}
This decoding rule retains a complexity advantage over a naive application of the generalized likelihood ratio test due to the structure of the code design \cite{cuvelier2021quantum}.

\begin{figure}[ht]
	\centering
	\includegraphics[scale = .47]{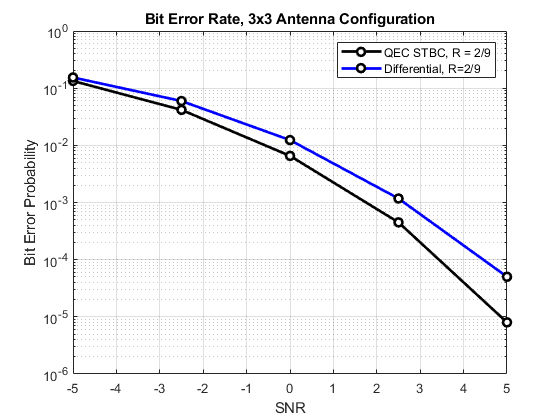}
    \vspace{-.3cm}
	\caption{Bit error rate comparison between differential code and quantum error correction based space-time block code (QEC STBC) for $3\times 3$, $d=3$, and a 4-packing for the codebook. The effective bit rate is $2/9$. We simulated 10 million channel realizations (assumed to be coherent for $T=9$ instances).}
	\label{fig:sq}
\end{figure}

\begin{figure}[ht]
	\centering
	\includegraphics[scale = .47]{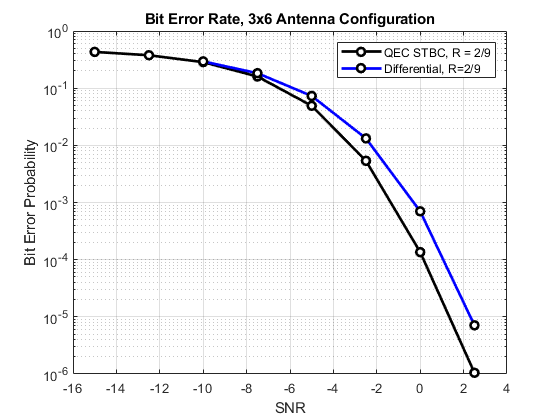}
    \vspace{-.3cm}
	\caption{Bit error rate comparison between differential code and quantum error correction based space-time block code (QEC STBC) for $3\times 6$ configuration,  $d=3$, and a 4-packing for the codebook. The effective bit rate is $2/9$. We simulated 10 million channel realizations (assumed to be coherent for $T=9$ instances).}
	\label{fig:nonsq}
\end{figure}

\subsection{Maximum-Likelihood Decoding for Non-Square Configurations}
Due to the possibility of independent block processing, the maximum-likelihood detection rule for non-square configurations is structurally similar to the square case, but successfully integrates the additional diversity provided by the additional receive antennas. In particular, the stabilizer code allows for recovery of $MN$ independently faded encoded vectors in the non-square configuration. This allows us to retrieve $MN$ signals $\vec{p}_z$ with a structure identical to that of Equation (\ref{eq:decoded_signal}), but now numbering $MN$ independent signals. The derivation can be carried through in the new dimensions and yields 
\begin{equation}
    \hat{\vec{s}}_{\textrm{ML, nsq}} = \textrm{arg}\max_{\vec{s} \in \mathcal{C}} \vec{s}^* \sum_{n = 0}^{MN - 1} \vec{p}_n \vec{p}_n^* \vec{s} \, .
\end{equation}
The only difference between the square and the non-square case is that more independently faded decoded vectors appear in the summation. This increases the diversity achieved by the code and reflects in the performance. 

\section{Conclusion} 
\label{sec:conc} 
In Figures \ref{fig:sq} and \ref{fig:nonsq} we assess the performance of the generalized codes for $3 \times 3$ and $3 \times 6$ antenna structures. The selected bit rate is low, at $2/9$, which can be interpreted as reflecting a low-rate but high-reliability requirement. The packing used is the best Grassmannian $4$-packing in $\mathbb{C}^3$ known to the authors, from \cite{lovepacks}. Higher rate communication is supported by using a denser packing, for which Grassmannian and equiangular packings are both suitable \cite{lovepacks, sloanepacks}. The performance of these codes is competitive compared to both coherent and noncoherent approaches in a rich scattering environment modeled by Rayleigh fading and additive white Gaussian noise. The curve slope in each case is commensurate with maximal diversity for the underlying antenna topology, providing high reliability in that environment. 

The extensions to the stabilizer-based space-time block code construction address the need for full-diversity code at variable rates and varying antenna configurations and coherence times. Since noncoherent communication hardware is typically resource limited, this new flexibility creates opportunities to effectively use communication hardware to improve reliability without channel estimation. The fundamental utility of the underlying encoding method and its robustness to noise is an interesting avenue for future investigation, as there are opportunities to explore tradeoffs between the dimension of the encoded vector, the density of the packing, and its eventual effect on the achievable rate. 


\bibliographystyle{IEEEtran}
\bibliography{qstbc} 

\end{document}